\documentclass[preprint,proceedings]{rmaa}

\SetYear{2003}
\SetConfTitle{Compact Binaries in the Galaxy and Beyond}
\title{Short-Term Radio--X-ray Correlations of Cygnus~X-1}

\author{
  T. Gleissner,\altaffilmark{1}
  J. Wilms,\altaffilmark{1,2}
  G. G. Pooley,\altaffilmark{3}
  M. A. Nowak,\altaffilmark{4}
  K. Pottschmidt,\altaffilmark{5,6}
  S. Markoff,\altaffilmark{4}\\
  M. Klein-Wolt,\altaffilmark{7}
  R. P. Fender,\altaffilmark{7}
  R. Staubert\altaffilmark{1}} 

\altaffiltext{1}{IAA T\"ubingen, Sand 1, 72076 T\"ubingen, Germany.}
\altaffiltext{2}{present address: Univ.\,of Warwick, Coventry, UK.}
\altaffiltext{3}{Astrophysics, Cavendish Laboratory, Cambridge, UK.}
\altaffiltext{4}{MIT, Center for Space Research, Cambridge, USA.}
\altaffiltext{5}{MPE, Garching, Germany.} 
\altaffiltext{6}{INTEGRAL Science Data Centre, Versoix, Switzerland.}
\altaffiltext{7}{University of Amsterdam, Amsterdam, the Netherlands.}

\suppressfulladdresses
\listofauthors{T. Gleissner,  J. Wilms, G. G. Pooley, M. A. Nowak,
  K. Pottschmidt, S. Markoff, M. Klein-Wolt,  R. P. Fender, R. Staubert}
\indexauthor{Gleissner, T.}
\indexauthor{Wilms, J.}
\indexauthor{Pooley, G. G.}
\indexauthor{Nowak, M. A.}
\indexauthor{Pottschmidt, K.}
\indexauthor{Markoff, S.}
\indexauthor{Klein-Wolt, M.}
\indexauthor{Fender, R. P.}
\indexauthor{Staubert, R.}

\begin{document}

\maketitle 

\boldabstract{We analyze simultaneous radio--X-ray data of Cygnus~X-1 from
  the Ryle telescope (RT) and \textit{RXTE} over more than 4\,a. We show that
  apparent correlations on short time scales in the lightcurves of Cyg~X-1
  are probably the coincidental outcome of white noise statistics.}

\boldabstract{spanish translation spanish translation spanish translation
  spanish translation spanish translation spanish translation spanish
  translation spanish translation spanish translation spanish translation
  spanish translation spanish translation spanish translation spanish
  translation spanish translation spanish translation spanish translation
  spanish translation}

As a measure of correlation between radio and X-ray emission, we calculate
the maximum cross-correlation coefficient, ccf, of simultaneous radio and
X-ray lightcurves, which are rebinned to a resolution of 32\,s and
smoothed. Every single X-ray lightcurve segment is cross-correlated with
the corresponding radio lightcurve, up to a maximum shift $\Delta t=\pm$10\,h.

GRS~1915+105 does show radio--X-ray correlations (Mirabel et~al.\@ 1998;
Klein-Wolt et~al.\@ 2002), qualifying it as a source to test the used ccf
procedure. We use a set of 120 simultaneous radio--X-ray lightcurves, taken
with the RT and \textit{RXTE}.

\begin{figure}[!t]
  \includegraphics[width=\columnwidth]{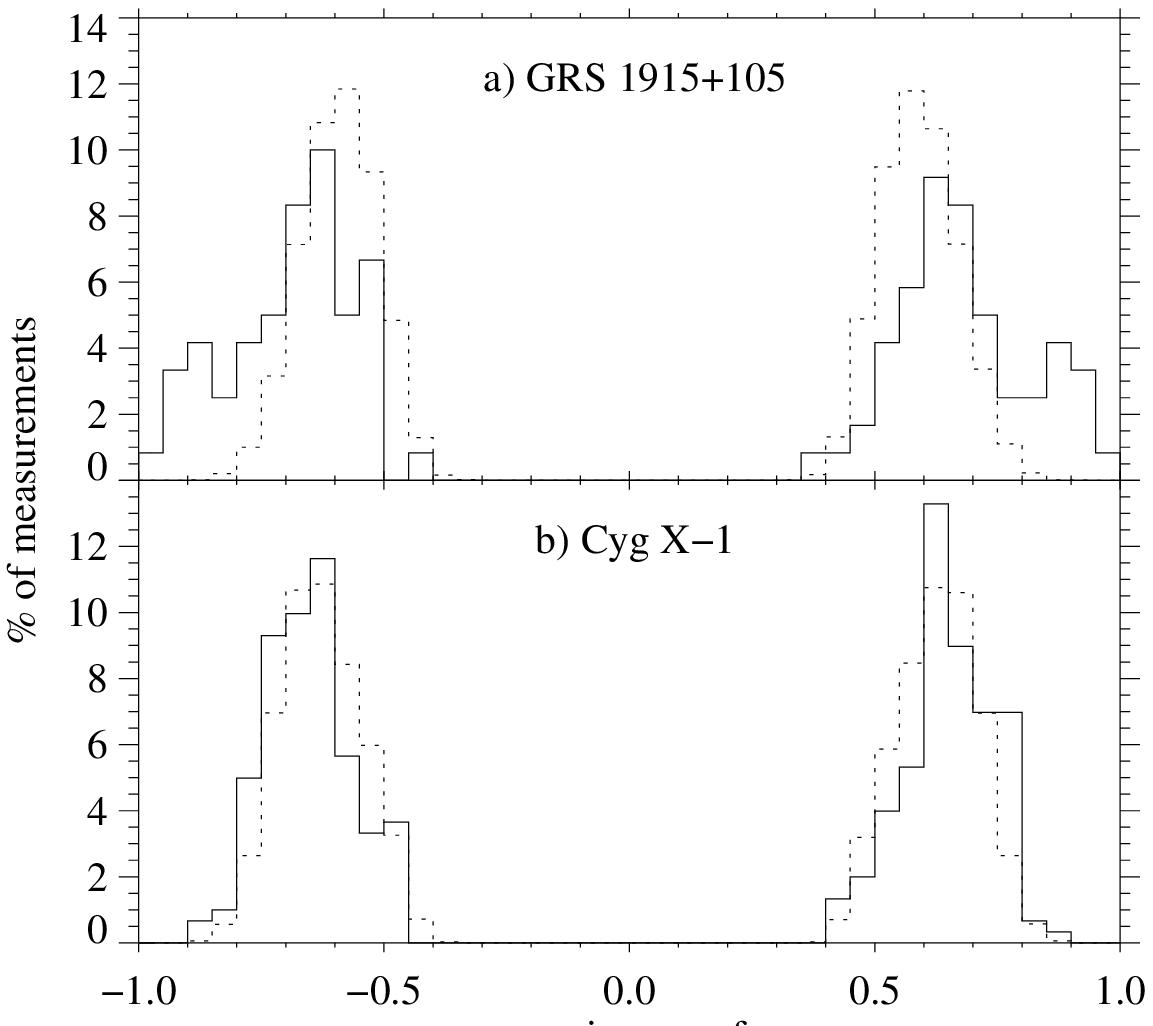}
  \caption{Histograms of the distribution of maximum ccf for observed
  data (solid line) and simulated wn data (dotted line) for GRS~1915+105
  (top) and Cyg~X-1 (bottom).}
  \label{fig:histogram}
\end{figure}

In Fig.~\ref{fig:histogram}a the histogram of the maximum ccf for the
observed data of GRS~1915+105 is drawn as solid line. The dotted line gives
the histogram of ccf from random white noise (wn) lightcurves which are
simulated with the same properties ($\mu$, $\sigma$, sampling) and
cross-correlated the same way as the observed lightcurves. For each
observation we run the simulation 1000 times to achieve statistical
significance.

Fig.~\ref{fig:histogram}a shows that the ccf distribution of the observed
data of GRS~1915+105 is significantly different from the wn distribution.
As expected, the existent radio--X-ray correlations in GRS~1915+105 are
reflected by the fraction of ccf's with values close to $\pm1$. This means
that the applied procedure is able to find radio--X-ray correlations in a
data set.

Applying the procedure to Cyg~X-1, we find that the ccf distribution of
301 observations is similar to the wn distribution
(Fig.~\ref{fig:histogram}b). This suggests that similar patterns detected
on short time scales in the X-ray and radio lightcurves of Cyg~X-1 are
random events which are a natural outcome in wn lightcurves. Taking into
account existent loose correlations on a time scale of days (Pooley
et~al.\@ 1999), this indicates a break-down of these correlations on
shorter time scales. One explanation is a sufficiently wide distance
between X-ray and radio emission regions so that small mass ejection
variations at the base of the jet are levelled out in the jet stream before
they reach the radio emission region. A detailed description will be given
in Gleissner et~al. (2004).

\end{document}